\documentclass[a4paper,11pt]{article}
\usepackage[utf8x]{inputenc}
\usepackage{epsfig,amssymb,amsmath,latexsym,color,pifont,graphicx,psfrag,abstract}
\newcommand{\p}{\varphi}
\newcommand{\om}{\omega}
\newcommand{\e}{\varepsilon}
\title{\textbf{Evolutionary design of non-frustrated networks of phase-repulsive oscillators}}
\author{Zoran Levnaji\'c\footnote{Correspondence and requests for materials should be addressed to Zoran Levnaji\'c (zoran.levnajic@fis.unm.si)} \\
Faculty of information studies in Novo mesto, Novo mesto, Slovenia}
\date{}

\begin{document}

\maketitle
\begin{abstract}
Evolutionary optimisation algorithm is employed to design networks of phase-repulsive oscillators that achieve an anti-phase synchronised state. By introducing the link frustration, the evolutionary process is implemented by rewiring the links with probability proportional to their frustration, until the final network displaying a unique non-frustrated dynamical state is reached. Resulting networks are bipartite and with zero clustering. In addition, the designed non-frustrated anti-phase synchronised networks display a clear topological scale. This contrasts usually studied cases of networks with phase-attractive dynamics, whose performance towards full synchronisation is typically enhanced by the presence of a topological hierarchy.\\ \\
\end{abstract}

\section{Introduction}
Study of collective phenomena in complex systems has been massively enhanced in the last decade by introduction of the \textit{complex networks} framework~\cite{dorogo}. This revolutionised our understanding of the real complex systems, ranging from gene and neuron networks, to air traffic, Internet and society~\cite{costa}. This paradigm allowed for novel theoretical and computational investigations, such as stability of gene regulatory networks examined through the network-inhibition of chaos~\cite{ja-bosa}, or detection of communities in social networks~\cite{fortunato}.

The phenomenon of synchronisation occupies a prominent place in the networks science, as a \textit{par excellence} example of emergent collective behaviour~\cite{prk-book}. Synchronisation of various kinds of oscillating dynamical units have been extensively studied both theoretically~\cite{arenas} and experimentally~\cite{motter}. The emergence of synchronisation is often examined by employing Kuramoto oscillators~\cite{k-book-a}. For being conceptually simple and easy to implement numerically, Kuramoto oscillators allowed for synchronisation to be examined in great detail~\cite{prk-book,arenas}. This model is often used in theoretical and experimental studies, such as network reconstruction~\cite{ja-arkady}, or modelling of neural phase-resetting curves~\cite{ramon}.

In parallel with synchronisation, research attention was devoted to various forms of anti-synchronisation~\cite{kim}. This particularly refers to anti-phase synchronisation~\cite{liu}, where the interacting oscillator pairs evolve towards having the opposite phases, i.e. towards being synchronised with a fixed phase difference $\pi$. In this context, networks of units with the repressive and/or repulsive interactions were investigated. Most famously, the system of repressively interacting genes -- \textit{repressilator} -- received much attention~\cite{e-g-h}. Repulsive interactions can generate \textit{dynamical frustration} and multistability, since interacting units cannot always relax to a unique equilibrium state~\cite{mogens,aneta}. Neurons are often coupled repulsively, which was examined theoretically~\cite{perc} and experimentally~\cite{gabor}. In addition, repulsive Kuramoto oscillators can be used to model social interactions and social equilibria~\cite{marvel-hong}, or suppression of synchronisation~\cite{louzada}.

Recently, complex networks of Kuramoto oscillators will solely repulsive coupling were examined~\cite{ja-sam}, in opposition to studies involving both attractive and repulsive interactions~\cite{zanette}. By borrowing the terminology from disordered systems~\cite{mezard}, the dynamical states in these networks were characterised by defining the \textit{link frustration} $f$. In general, complex networks were found to have positive total frustration, due to complexity of their topologies. In addition, depending on the initial conditions, networks typically display multiple final dynamical states. The network's emergent dynamics is thus strongly dependent on the topology, and for a general network does not reach ``minimum energy'' state with zero frustration.

In a different context, researchers have been studying the problem of network design, motivated by the technological need to engineer networks that perform prescribed functions. Models are typically based on a network whose topology is evolving towards the optimal target topology. Most used among such models is the evolutionary optimisation, a version of Monte Carlo simulated annealing algorithm~\cite{bornholdt}. A myriad of results were obtained through this formalism, such as emphasising peculiarities of the scale-free topologies~\cite{panos}, design of biological transduction networks~\cite{kaluza1,kaluza2}, design of robust genetic clocks~\cite{kobayashi}, or design of networks that optimally perform a given dynamical behaviour~\cite{karalus}. In the context of oscillators, design of easily synchronisable network was recently examined~\cite{yanagita}, and a method of achieving the prescribed synchronisation state by configuring a repulsive subnetwork was exposed~\cite{arizmendi}.

A question that remains open in the context of phase-repulsive networks, regards the topologies that allow for the zero frustration state to be attained. These are the networks whose dynamics leads to anti-phase synchronisation for any choice of initial conditions. While essentially any oscillator network with phase-attractive coupling generates full synchrony, obtaining anti-phase synchronisation is more demanding. As shown in~\cite{ja-sam}, despite that phase-repulsive interactions naturally tend towards anti-phase synchrony, a generic network typically can not attain fully anti-phase synchronised state due to the topological constraints. In this paper we employ the evolutionary design algorithm~\cite{bornholdt,kaluza1,kobayashi}, to obtain phase-repulsive oscillator networks that always achieve non-frustrated, i.e. anti-phase synchronised state. By choosing the links to rewire according to their frustration, the initial network topology evolves towards minimising global frustration, which inevitably leads to non-frustrated topology. As we show, in opposition to the well studied synchronisable networks~\cite{yanagita}, whose synchronisation performance is typically enhanced by hierarchical or scale-free structures~\cite{kurths}, non-frustrated networks are topologically very uniform, and exhibit a precise scale~\cite{tanaka,natasa} rather than being scale-free. This problem is similar to the problem of 2-color vertex colouring encountered in graph theory~\cite{riste-diestel}. The difference is that the process of vertex colouring is here automatically done by the network phase-repulsive dynamics.

We develop our model by considering a network that consists of $N$ oscillators (nodes), connected via $L$ non-directed links. Dynamical state of the oscillator $i$ is given by the phase variable $\p_i \in [0,2\pi)$, and its dynamics is defined by:
\begin{equation}  \dot{\p_i} = \om_i + \frac{\e}{k_i} \sum_{j=1,N}  A_{ij}  g(\p_j - \p_i) \; ,  \label{eq-1} \end{equation}
where $k_i$ is the node's degree ($\sum_i k_i = 2L$), $\e$ is the coupling strength, and $\om_i$ is the oscillator's natural frequency. Network's topology is expressed via symmetric adjacency matrix $A_{ij}$, defined as $A_{ij}=1$ if nodes $i$ and $j$ are connected, and $A_{ij}=0$ otherwise. We consider identical oscillators $\om_i = \om$, and set $g=\sin$, reducing our system to the simple Kuramoto model~\cite{k-book-a}. Following the results exposed in \cite{ja-sam}, we examine the network model with phase-repulsive coupling. To that end we fix $\e=-1$, and for simplicity take $\om = 0$. The Eq.\ref{eq-1} for our model reads:
\begin{equation} \dot{\p_i} =  -\frac{1}{k_i} \sum_{j=1,N} A_{ij} \sin (\p_j - \p_i) \; . \label{eq-2} \end{equation}
Time-evolution starts from a random set of initial phases (IP), selected independently for each oscillator from $\p_i (0) \in [0,2\pi)$. The interacting oscillator pairs evolve towards maximising the phase difference between them, i.e. their phase values ``stretch'' apart from each other as much as possible. Preferably, the maximal phase difference is $\pi$. However, for a general network, due to its complex topology, the phase difference along certain links is often less than $\pi$, or even zero. The final (stationary) dynamical state is quantified by introducing \textit{frustration} $f_{ij}$ for each link as~\cite{ja-sam,zanette}:
\begin{equation} f_{ij} = A_{ij} [ 1 + \cos (\p_j - \p_i) ] \; . \label{eq-f} \end{equation}
Frustration measures how ``squeezed'' is a link: it can be pictured as the elastic potential energy contained in it: a link stretched to the phase difference $\pi$ has zero frustration, whereas a link forced to synchronise carries the maximal frustration that equals 2. We characterise the network dynamical states by specifying the link frustration values $f_{ij}$. To measure the global frustration we define $F$ as the network average of $f$:
\begin{equation} F = \frac{1}{L} \sum_{i>j} f_{ij} \; , \label{eq-F} \end{equation}
which quantifies how much does the network topology allow for links to stretch. Note that $F$ can be seen as the non-equilibrium potential, since Eq.\ref{eq-2} can be written as~\cite{zanette}:
\begin{equation}  \dot{\p_i} =  - \frac{2 L}{k_i} \frac{\partial F}{\partial \p_i} \; . \end{equation}
As concluded in \cite{ja-sam}, phase-repulsive dynamics for a general network leads to a positive total frustration, whose value strongly depends on the details of network topology. In widely studied phase-attractive case, full synchronisation is essentially the only final dynamical state, independently of the topology. In contrast, topology in our model crucially determines the dynamics. In addition, a general phase-repulsive network displays multiple final frustration states, each occurring for a certain fraction of IP.

Interestingly, some networks exhibit a unique dynamical state with zero total frustration $F=0$. These networks are completely anti-phase synchronised -- each coupled oscillator pair has the opposite phases for any IP. This comes from peculiar topologies of such networks that allow for all links to stretch, thus completely avoiding frustration. While in the phase-attractive case, basically any network achieves the fully synchronised state, in the phase-repulsive case only a limited set of networks attain the anti-phase synchronised state. In addition, such state is also stable, as it is the only possible dynamical equilibrium state. How can we construct non-frustrated networks? What are their topological properties? The rest of this paper is devoted to these two questions.


\section{Results}
We employ the evolutionary algorithm based on simulated annealing in order to design non-frustrated networks. The evolutionary process involves rewiring links with the probability proportional to their frustration $f_{ij}$, until a network with zero frustration along all links for all IP is obtained. Further details on our network design method are provided in Methods section below.

\begin{figure}[!ht] \centering
  \includegraphics[width=0.3\textwidth]{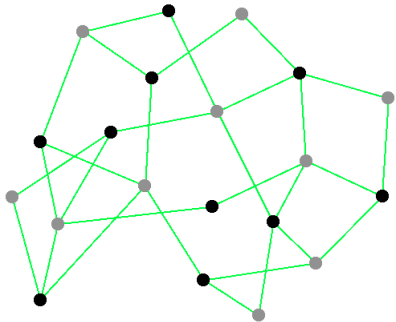}
  \caption{Example of an evolutionarily designed non-frustrated network, with $N=20$ nodes and $L=30$ links. All adjacent node pairs are having the opposite phase values (phase difference $\pi$), and hence zero frustration, for any choice of IP. Nodes appear in two shades in order to show bipartitness. Each node has zero clustering.} \label{newfig-1}
\end{figure}

An example of a simple evolutionarily designed non-frustrated network with $N=20$ nodes and $L=30$ links is shown in Fig.\ref{newfig-1}. It was obtained after 15 evolution steps. 
The network appears to be uniform in structure, without prominent hubs. It is also bipartite, as illustrated by different shades of adjacent nodes. Large presence of 4-node rings, which are small non-frustrated networks, is clearly visible. There are also 6-node rings, which are unable to exhibit the higher $F$ state due to being topologically intertwined with the rest of the network, which is forcing them into $F=0$ state. Clustering coefficient is zero for all nodes. Due to a large number of 4-node rings, the network topology seems more regular than expected from an evolution governed by random rewirings.

\begin{figure}[!ht] \centering
  \includegraphics[width=\textwidth]{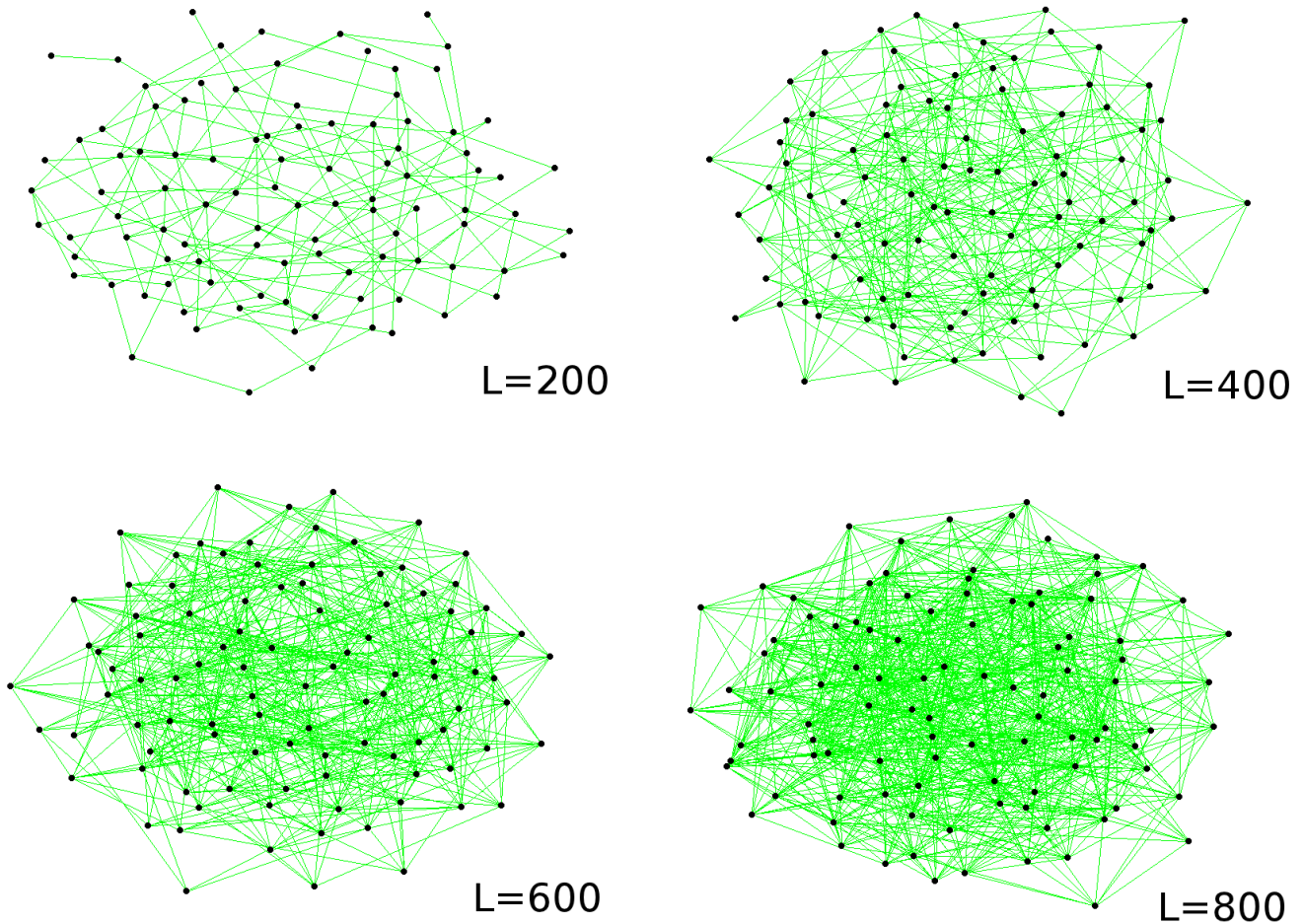} 
  \caption{Four networks with zero total frustration, obtained through evolutionary processes described in the Methods section, with $N=100$ and $L$ as indicated. Again, networks attain anti-phase synchronisation for any choice of IP, are bipartite and have zero clustering. Networks of size beyond 800 are discussed in the Supplement figure.} \label{newfig-2}
\end{figure}

To quantify more precisely the topological properties of non-frustrated networks, we now consider larger networks with $N=100$ nodes. We run evolution simulations with various numbers of links, ranging from $L=200$ to $L=2000$: all evolution processes yield a final non-frustrated network. In particular, we here focus on networks with $L=200, 400, 600$ and 800, which are shown in Fig.\ref{newfig-2}, and indicated via the corresponding $L$ (details on this evolution process are included in the Methods section below).
As observed earlier with Fig.\ref{newfig-1}, the non-frustrated networks seem not to posses any hierarchical structure, which is even more apparent here. Of course, all the obtained networks are bipartite and have zero clustering. We run many evolution simulations with the same $N$ and $L$: the same global topological features were always observed. As we show more quantitatively in what follows, the structures of these networks are indeed surprisingly regular.

\begin{figure}[!ht] \centering
  \includegraphics[width=0.5\textwidth]{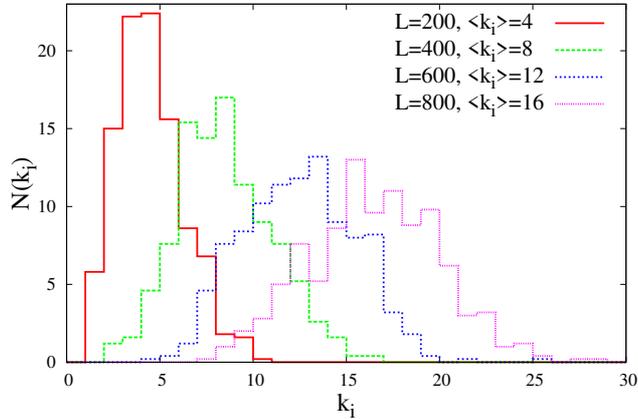} 
  \caption{Histograms of node degrees for four non-frustrated networks with $N=100$ and $L$ as indicated, in correspondence with Fig.\ref{newfig-2}. Data is averaged over 5 realisations of such networks. Each histogram displays a prominent peak, coinciding with the average node degree for each network.} \label{newfig-3}
\end{figure}

We first compute the histograms of node degrees. To this end, we average the degree histograms over 5 realisations of the non-frustrated networks with given $N$ and $L$, and report the results in Fig.\ref{newfig-3}. All histograms are very localised in range and display a well pronounced peak in the centre, whose value coincides with the average degree $\langle k_i \rangle = 2 \frac{L}{N}$. 
As expected, for the increasing network size, the peak becomes gradually less pronounced. This confirms that our networks are not hierarchical. In order to nullify the frustration, network evolution leads towards evenly spreading the remaining link frustration over the network. This indicates that rather than being scale-free, non-frustrated networks are ``scale-rich'', i.e. posses a well defined scale~\cite{tanaka} (this holds for small $L$, i.e. sparse networks). The scale is determined by the average node degree.

The node degree histograms suggest that non-frustrated networks posses structures that are similar to Erd\H os-R\'enyi random networks, that the evolutionary process started from. However, two types of structure also have profound differences. Non-frustrated networks, by being bipartite, always have zero clustering. In contrast, the average expected clustering coefficient for random networks is $\frac{\langle k_i \rangle}{N}$~\cite{dorogo}. This difference is a clearly pronounced in dense non-frustrated networks with high $\langle k_i \rangle$. In addition, we found that non-frustrated networks systematically display somewhat longer average shortest paths than random networks with the same $N$ and $L$. This can again be attributed to zero clustering: since no three nodes can be mutually interconnected, shortest paths are on average somewhat longer. Regardless of $N$ and $L$, average shortest path in any non-frustrated network can never exceed $\frac{3}{2}$, since on average, for each node, a half of the remaining nodes are at least two links away.

There is an upper bound on $L$ for any given $N$, beyond which it is not possible to design non-frustrated networks, since avoiding connected node triangles becomes impossible. Close to this limit the network evolution process gradually becomes slower. The network structure however becomes even more regular, since avoiding frustration requires more and more ordered organisation of links, and more uniform node degrees. In the supplementary figure to this paper, we show the additional six networks with $N=100$, and $L$ ranging from 1000 to 2000. The regularity and bipartitness of these dense networks are clearly visible.

\begin{figure}[!ht] \centering
  \includegraphics[width=0.5\textwidth]{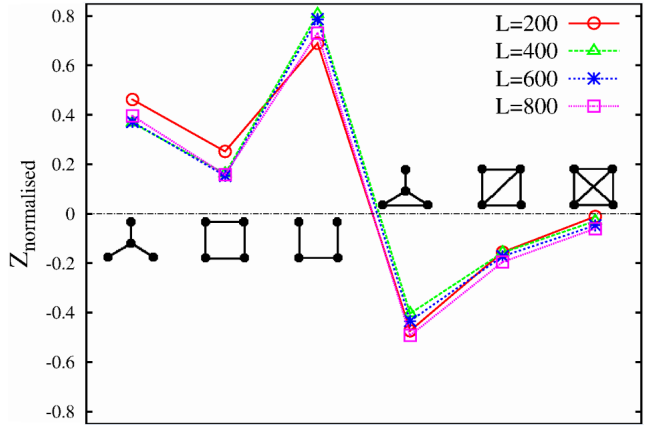} 
  \caption{The value of $Z_{normalised}$ for all non-directed 4-node motifs in the networks considered in Fig.\ref{newfig-2} as indicated, computed using \textit{mfinder} software~\cite{mfinder}. Non-frustrated motifs are over-represented, while the frustrated motifs are under-represented. Profiles are invariant to the network size.} \label{newfig-4}
\end{figure}

Finally, we quantity the topological regularity of non-frustrated networks. We achieve this by measuring the concentration of motifs in our networks, using the \textit{mfinder} software~\cite{mfinder}. We first introduce $Z_{score}$, as a measure of how over- or under- represented is a given motif $w$ within some network. It is defined as~\cite{milo}: 
\begin{equation}  Z_{score}(w) = \frac{N_{actual}(w) - N_{random}(w)}{SD(w)} \; ,  \end{equation}
where $N_{actual}(w)$ is the number of motifs $w$ found in the examined network, $N_{random}(w)$ is the average expected number of those motifs in a network with the same $N$ and $L$, while $SD(w)$ is the standard deviation of $N_{random}(w)$. Thus, any motif whose $Z_{score}$ is considerably bigger that 1 (-1) is over- (under-) represented. Since the studied networks differ in their numbers of links, rather them employing $Z_{score}$, we instead rely on normalised $Z_{score}$, which is defined as~\cite{kaluza2}:
\begin{equation}  Z_{normalised}(w) = \frac{Z_{score}(w)}{\sqrt{ \sum_w Z_{score}(w)^2 }} \; .  \end{equation}
We calculate $Z_{normalised}$ for all 4-node non-directed motifs (shown in Fig.\ref{newfig-5}c), for four evolutionarily designed networks shown in Fig.\ref{newfig-2}. The results are reported in Fig.\ref{newfig-4}.
In all networks, non-frustrated motifs are overwhelmingly over-represented, while the frustrated motifs are drastically under-represented (cf. Fig.\ref{newfig-5}). Both over- and under- representation are invariant to varying realisations of non-frustrated networks, and to the number of links $L$. The same result is found for both 3-node motifs: the chain is systematically over-represented, while the ring is strongly under-represented. This suggests that, regardless of their size, non-frustrated networks have an inherent motif pattern, rooted in their dynamical properties. This result confirms that large non-frustrated networks are essentially composed of small non-frustrated motifs. By locally forming non-frustrated motifs, mechanism of network evolution evolves towards global zero frustration. Interestingly, this means that the emergence of global $F=0$ is achieved locally, by each network building block evolving towards $F=0$. This implies that any large non-frustrated network can be decomposed into smaller non-frustrated subnetworks (an alternative scenario can be imagined, where the desired global property is achieved only globally, i.e. only at the level of the entire network, with no subnetworks exhibiting the same property).


\section{Discussion}
Natural evolutionary processes in real complex networks often lead to scale-free structures in the emerging networks~\cite{panos}. However, the ubiquity of scale-free topologies has been recently challenged. New findings include examples of real complex networks, that despite emerging from biological self-organisation, posses a clear scale in their structure, contrasting widely accepted universality of scale-free topologies~\cite{tanaka,natasa}. In this paper, we showed a model of network topology that develops under a simple evolutionary constraint, whose topology evolves away from hierarchical structure, and towards a ``scale-rich'' structure~\cite{tanaka}. As shown above, this can be attributed to the nature of phase-repulsive dynamics, which requires certain topological properties in order to attain fully anti-phase synchronised state (i.e., non-frustrated state). Hence, in opposition to the phase-attractive dynamics and resulting full synchronisation, that are typically improved by the presence of a topological hierarchy, anti-phase synchronisation appears to be enhanced by the lack of hierarchy.

Bipartite networks are usually considered in relation to two different types of nodes, such as on-line comments and the corresponding web-page users (persons)~\cite{marija}. Here we showed that bipartitness can spontaneously emerge from the evolution of a network with a single type of nodes (identical oscillators). Besides these findings, this work confirmed again that evolutionary design through random link rewirings is a viable method of constructing complex networks that exhibit prescribed dynamical patterns. Our method mimics the natural evolutionary processes, by implementing random mutations (rewirings). To make mutations completely random, one could modify the algorithm to choose links for rewiring regardless of their frustration. This would however only slow down the process, without changing the final result. On the other hand, it is possible that the non-frustrated networks can be obtained by means of other design methods. Such methods are however still poorly explored, even in the context of simple models such as oscillator networks.

We conclude this section by examining the limitations and extensions of our results. As noted earlier, zero frustration is not achievable beyond a certain limit of $L$. Our algorithm can in this case be applied to minimising the frustration, and studying the ways in which the resulting networks organise the remaining frustrated links. Another question refers to maximising instead of minimising the total network frustration. By accepting the mutations that increase rather than decrease $F$, our algorithm can be used to design networks with maximal obtainable frustration. Intriguing problem here is the difference between the frustrated (squeezed) and non-frustrated (relaxed) network topologies. This is related to elastic network models, where mechanical properties of networks of elastic springs are studied~\cite{holger}. It would be also interesting to apply the same algorithm to design non-frustrated networks using other models of repulsive interactions. This particularly refers to neural and gene interactions, where networks with repressive interactions are widely studied~\cite{mogens,aneta,perc}. This could aid engineering genetic or neural networks with a given prescribed functionality. Finally, we note that our networks were designed to satisfy a simple dynamical rule, formulated as $F=0$. By varying the dynamical rules, one could use this or similar algorithms to design networks whose topological properties reflect a predetermined dynamical rule. This could lead to the discovery of new complex network topologies, that do not fit into any of the currently studied topological categories.


\section{Methods}

In this section we describe our method of evolutionarily network design and explain its implementation. \\

\begin{figure}[!ht] \centering
  \includegraphics[width=0.4\textwidth]{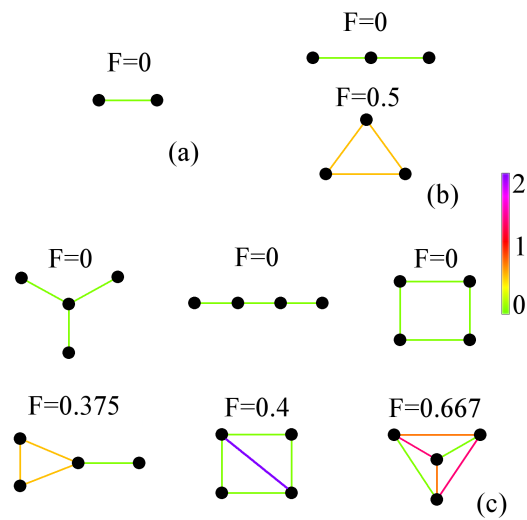} 
  \caption{Frustration for two-node network in (a), both three-node networks in (b), and all six possible four-node networks in (c). Total frustration $F$ is reported, along with all link frustrations $f_{ij} \in [0,2]$, illustrated by the (color)scale. Frustration is defined in Eq.\ref{eq-f} and Eq.\ref{eq-F}, and refers to the dynamical model Eq.\ref{eq-2}. Note that three four-node networks are frustrated, and the other three are not.} \label{newfig-5}
\end{figure}

\begin{figure}[!ht] \centering
  \includegraphics[width=0.4\textwidth]{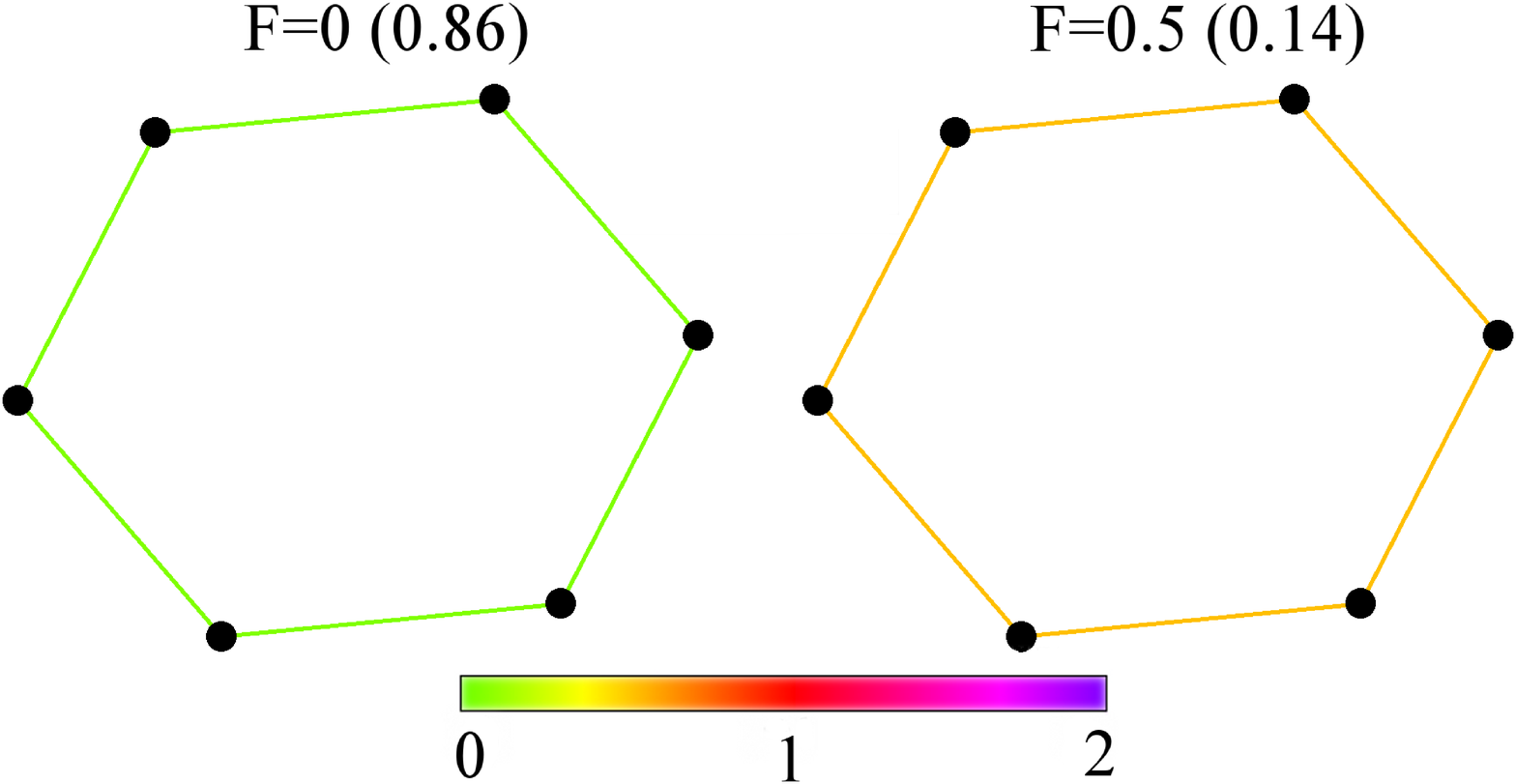} 
  \caption{Two frustration states exhibited by the 6-ring. Total frustration $F$ and the link frustrations $f_{ij} \in [0,2]$ are shown for each state. The non-frustrated and the frustrated states appear for 86\% and 14\% of IP, respectively.} \label{newfig-6}
\end{figure}

\noindent \textbf{Preliminaries.} For small networks it is easy to find non-frustrated topologies, as shown in Fig.\ref{newfig-5}. A connected pair of nodes, 3-node chain, 4-node star, 4-node chain and 4-node ring (cycle), are networks that show a single frustration state $F=0$.
However, finding a large non-frustrated network of size $N \gg 1$, is not straightforward. A typical network will in general have $F>0$, for instance, due to having at least one 3-node ring. On the other hand, it is easy to notice that for a network to be non-frustrated, it needs to be bipartite, and hence have zero clustering. Indeed, for having two groups of nodes with opposite phases, the network needs to involve links that exclusively go from nodes of one group to the nodes of the other group. However, although necessary for anti-phase synchronisation, bipartitness is not sufficient for it. In Fig.\ref{newfig-6} we show the 6-node ring, which is a bipartite network. However, besides displaying the expected non-frustrated state, the network possesses another frustrated state with $F=0.5$, that occurs for 14\% of IP.
Bipartitness alone guarantees the existence of a non-frustrated state, but it does not guarantee its uniqueness. This indicates that we need a more elaborated mechanism for designing non-frustrated networks with the \textit{unique} $F=0$ state. \\

\noindent \textbf{Evolutionary algorithm.} We employ the simulated annealing algorithm as follows. Starting from a random network, at each evolution step we rewire one of the frustrated links, and accept the mutation if the new network exhibits smaller total frustration $F$ than the original one. The network topology evolves until the target $F=0$ is reached. More precisely, our algorithm goes as follows:
\begin{description}
\item[0.] Start from a non-directed, connected network ${\mathcal N}$, with $N$ nodes and $L$ links ($N-1 \le L \le \frac{N(N-1)}{2}$), and no self-loops. We use Erd\H os-R\'enyi random network. Compute link frustration values $\bar f_{ij}$ and total frustration $\bar F$, averaged over $M$ random IP. We use $M=5$.
\item[1.] Select a link $ij$ in ${\mathcal N}$ with probability proportional to $\bar f_{ij} + \alpha$ (we use $\alpha=0.01$). Detach this link from one of its end-nodes (randomly chosen), and rewire it to a different node chosen with uniform probability, making sure that the network stays connected and without self-loops. This is the mutated network ${\mathcal N}'$.
\item[2.] For ${\mathcal N}'$ compute $\bar f_{ij}'$ and $\bar F'$. If $\bar F' < \bar F$ the mutation is accepted, and the network is updated: ${\mathcal N} \equiv {\mathcal N}'$. If $\bar F' > \bar F$ the mutation is accepted with probability $e^{-(\bar F' - \bar F)/ \sigma \bar F}$. Otherwise, the mutation is rejected and ${\mathcal N}$ remains unchanged. We use $\sigma=0.02$.
\item[3.] The process continues from step 1 by selecting a new link in ${\mathcal N}$ to be rewired. 
\end{description}
The evolution continues until the final network ${\mathcal N}^0$ with a unique frustration state $F=0$ is reached. Potential multiple frustration states are excluded by averaging over $M$ random IP. The numerical constant $\alpha$ skews the probability distribution allowing for non-frustrated links to be mutated as well, while ensuring that more frustrated links are still mutated more often. This drastically expedites the evolution process, while at the same time allowing for all links to be mutated. Parameter $\sigma$ allows that, with a small probability, even mutations leading to a small increase of total frustration get accepted. By playing the role of temperature, $\sigma$ prevents the evolution process from getting stuck into local minima.\\

\begin{figure}[!ht] \centering
  \includegraphics[width=0.4\textwidth]{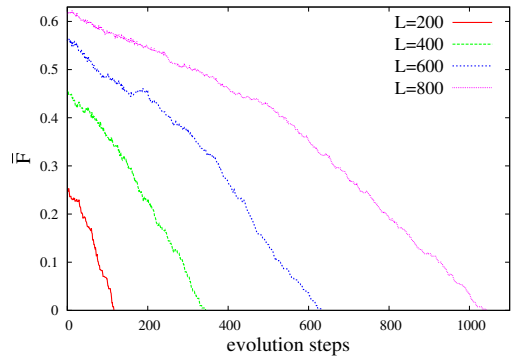} 
  \caption{Network evolution processes generating four networks considered in Fig.\ref{newfig-2}. The value of total frustration averaged over $M=5$ IP ${\bar F}$ is shown, as function of the number of evolution steps.} \label{newfig-7}
\end{figure}

\noindent \textbf{Algorithm's performance.} We tested the evolution algorithm exposed above on numerous network examples: the process invariably leads to a non-frustrated network. Since the evolution involves randomness, the final network does not depend on the starting one. On the other hand, for any fixed $N$ and $L$ there are many different realisations of final non-frustrated networks, all sharing the key topological properties. To illustrate the progress of the network evolution, we show in Fig.\ref{newfig-7} the gradual decrease of $\bar F$ for four examples of evolution simulations related to networks from Fig.\ref{newfig-2}.
The evolution speed decreases with the increase of $L$. In a denser network it is harder to find a mutation that leads to a decrease of $\bar F$, and when found, that mutation typically decreases $\bar F$ for a small amount. In addition, denser networks start with a larger initial $\bar F$, thus additionally increasing the needed number of evolution steps. We adjusted for the optimality of the process through the parameters $M$, $\alpha$ and $\sigma$. The ``bumpiness'' of the processes is due to Monte Carlo algorithm's temperature $\sigma>0$. This parameter needs to be adequately set to optimise the evolution speed and effectiveness. Too small $\sigma$ might result in process getting suck into a local minimum of $\bar F$, which may incorrectly suggest that smaller $\bar F$ is not obtainable. Too large $\sigma$ could induce the process to wander too excessively, without ever finding $\bar F = 0$ solution. Our choice of $\sigma=0.02$ takes into account both issues. At each evolutionary iteration, the network usually has multiple frustration states with different $F$-values. To obtain a single frustration value $\bar F$ corresponding to the current network, we average $F$-values over $M=5$ IP. This is an additional reason for introducing temperature $\sigma$: due to a potentially wide range of multiple states, averaging might not give the best measure of the total frustration. Upon each completion of evolutionary simulation, we checked that the obtained network is indeed non-frustrated.


\noindent \textbf{Acknowledgments.} Thanks to Alexander Mikhailov for useful suggestions, and Uri Alon for making \textit{mfinder} publicly available.\\\\

\noindent \textbf{Author Contribution.} Zoran Levnaji\'c has done the entire work related to this paper.\\\\

\noindent \textbf{Competing Financial Interests.} The author declares no competing financial interests.\\\\

\noindent \textbf{Figure Legends:}
\begin{description}
\item[Figure 1] Simple example of an evolutionarily designed non-frustrated network.
\item[Figure 2] Large evolutionarily designed non-frustrated networks.
\item[Figure 3] Histograms of the node degrees in non-frustrated networks.
\item[Figure 4] Motif concentrations in non-frustrated networks.
\item[Figure 5] Frustration in small networks.
\item[Figure 6] Two frustration states exhibited by the 6-ring network.
\item[Figure 7] Illustration of the network evolution processes.
\end{description}

\end{document}